# TRANSCENDING OLD BOUNDARIES: DIGITAL AFTERLIFE IN THE AGE OF COVID-19

Mashiat Mostafa, RMIT University, mashiat.mostafa@rmit.edu.au

Faheem Hussain, Arizona State University (ASU), faheem.hussain@asu.edu

**Abstract:** The primary objective of our exploratory research is to contribute to the ongoing conversation on Digital Afterlife from the lenses of Global South during the COVID-19 period. Digital Afterlife is fast becoming a challenge for our increasingly connected society. Moreover, the situation got worse with the COVID-19 pandemic. The on-going research is to address the disparity in the Global South, specifically in countries like Indonesia, India and The Philippines compared to the Global North for Digital Afterlife services such as policies and digital mourning services. By addressing the research question, 'What services and policy frameworks are available for Digital Afterlife in the Global South during COVID-19?', we aim to find the multitude of ways people in the Global South are managing their digital footprints. Our preliminary findings show that some considerable research and death related digital services and innovation have taken place during the pandemic. However, overwhelming majority of these works are western-centric and mainly dealing with post-mortem personal asset management. Cultural nuances, socio-economic perspectives, religion, political climate, regional infrastructures are mostly sidelined. We found significant disparity in Digital Afterlife product and service designs, which got worse during the global pandemic. Our goal is to collect further in-depth data within the three big ICT powerhouses of global south (Indonesia, India and The Philippines), identify the challenges as well as the innovations around Digital Afterlife. We envision proposing a set of recommendations, based on our findings, for developing a more inclusive and equitable digital space in this pandemic-stricken world.

## 1. INTRODUCTION

1. This research aims to develop an in-depth understanding whether various online services foster digital dependencies in the Global South (developing nations) and Global North (developed nations) specially for Digital Afterlife services (Dados & Connell, 2012; Lemuel, 2010). This is an ongoing research to understand the ways different technology services have affected the Global South society in dealing with death and related consequences specifically in the most densely populated and COVID-19 affected countries like, Indonesia, India and The Philippines (World Meter, 2020). This knowledge can help us to design, develop, and manage inclusive and efficient technological solutions for Digital mourning services with social and religious empathy in the COVID-19 world. Our previous research shows that, there is a considerable gap in the Digital Afterlife scholarship for addressing the developing region's (Global South) (Hussain, et.al, 2017). Moreover, due to COVID-19 it is now significant that we recognize the gaps and challenges faced in the Global South for both Digital Afterlife and Digital Mourning services.

2. Connections with numerous digital services (Facebook, Twitter and Instagram) have ensured better access to information, citizen services, communication, productivity, and much more.

However, such a digitization process enforces every user to leave a part of their presence, big or small, in digital formats in cyberspace. Digital Afterlife is the digital footprint we leave behind through our Google mails, Facebook accounts, Amazon's online services, or Apple's iTunes (Hussain, et.al, 2017). After our biological death, these digital remnants would continue to exist in cyberspace, as our Digital Afterlife. Information and Communication Technologies (ICTs), as we know, are yet to be designed to properly manage the eventual death of their users and the related inherent challenges.

3.   Based on a rapid and in-depth qualitative ethnographic research, we seek to answer the following questions:

I. What services are available for Digital Afterlife in the Global South during COVID-19?

   a) Will digital mourning services become mainstream COVID-19? How will this affect religious, cultural and traditional ceremonies?
   
   b) Are there localised and customised services for the focused countries in this research?

II. Are there policy frameworks available to address Digital Afterlife in Global South?

   a) Are all the global services ensuring same or equivalent policies for both Global South and Global North?
   
   b) Are there any new policies for Digital Afterlife in India, Indonesia and The Philippines during COVID-19?

## 2. Digital Afterlife: Current Practices and Policies

According to HCI, Kaye et al. talked about "legacy" elements of any user, which represent an individual's digital work (Kaye, et. Al, 2006). Massimi and Charise mentioned the need of sensitive orientation of any technological solution in dealing with the deceased's data (Brubaker, 2015; Massimi & Charise, 2009). ICT providers are yet to be designed to effectively acknowledge and reflect the eventual death of their users (Vitak; et.al., 2012). Nevertheless, with few exceptions like the works of Gray and Paul, who recognized the need of inclusion and diversity in design, majority of the works on digital post-mortem literature had an underlying assumption of somewhat universality of Digital Afterlife related challenges (Ellis & Coulton, 2013). The cultural nuances, socio-economic perspectives, political climate, regional infrastructure- the factors which have proven to be immensely critical for ICT diffusion in Global South are not considered and included in the present Digital Afterlife scholarship.

Based on the lack of awareness and understanding about Digital Afterlife, people are confused with which laws or social practices to follow as far as the digital bequeathing is concerned. Up until now, there has been no consensus on following a common set of policies in this regard. The Uniform Law Commission (ULC) in United States approved the Uniform Fiduciary Access to Digital Assets Act RUFADAA in 2015 (Uniform Law Commission, 2015). The RUFADAA does not allow any decedent or incapacitated person's digital assets to be disclosed to anyone, even to their fiduciary. In such situations, the digital service providers will have control of any deceased users' digital assets. Until now, at least 46 states have enacted laws addressing access to email, social media accounts, or certain electronically stored information, upon a person's incapacity or death (Uniform Law Commission, 2015). In contrast, European Union (EU) leaves discretion in implementation to EU member states to extend this minimum protection, which is guaranteed. Some European Union countries have used this possibility, and their data protection laws offer post-mortem data protection, limited in its scope and post-mortem duration (Harbinja, 2013).

ICT services, online and offline in the Global South are not designed to include options for dead users or the management of digital resources after the corresponding users' death. The major net based global entities like Facebook, Google, Twitter are now putting considerable provisions, and policies in place to deal with Digital Afterlife. "Firms such as Eterni.me and Replica now offer consumers online chat bots, based on one's digital footprint, which continue to live on after users die, enabling the bereaved to 'stay in touch' with the deceased" (Öhman & Floridi, 2018).

During COVID-19 quite a few services were introduced in the Global North. There are four categories of firms in the Digital Afterlife Information - Information management services, posthumous messaging services, online memorial services, re-creation services. One can now create wills online and download them to a Digital Afterlife service provider. Along with that some Digital Afterlife service providers offer free data management and others require a once off fee or monthly/yearly subscription to remain in use (Racine et al, 2020). In April 2020, funeral live streaming start-up, OneRoom, was working to meet demand to install cameras into funeral homes to offer funeral streaming services (Kuri, 2021).). Sadly, these new options are too western society specific and do not include Global South oriented challenges such as absence of copyright laws, scarcity of localized ICT services, poor governance, apathy of policymakers and service providers, etc. Moreover, such services are based or priced at a rate that is not feasible or intended for the standard of living in many developing countries, specifically low per capita income countries like India, Indonesia and The Philippines (Dados & Connell, 2012). Given the disparity in living standards these new Digital Afterlife services are becoming farfetched in many parts of the world due to lack of awareness. Even though COVID-19 has raised more demand for such Digital Afterlife services however, the lack of inclusion, cultural awareness resulted in limited or no services for the Global South.

## 3. RESEARCH METHOD

4.     This qualitative research has two parts. The first half is primarily based on in-depth literature review and ICT policy and service provision analyses. Major digital service providers with global footprints, prominent Digital Afterlife specific legal provisions, and real-world case studies are the main sources for our research, alongside key academic research on this interdisciplinary topic.

At the second phase of this research, we plan to engage with topic experts across different regions through in-person interviews and focus group discussions. Initially, we aim to work with approximately 20 to 50 respondents in each of our country sites in India, Indonesia, and the Philippines. The number of research participants will increase after that. For this research, online interviews will be conducted with religious, legal, social, and technical experts. Approximately 4 to 5 experts will be interviewed individually via online platforms. In addition, online interviews and focus group discussions with people who recently have availed different Digital Afterlife options for themselves or their deceased family members or friends; and with people who have not yet used any Digital Afterlife options for any of their digital accounts online.

## 4. Initial Findings: Emerging Challenges of Digital Afterlife in the Time of COVID-19

### 4.1. Increasing Digital Legacies Without Proper Guidelines

At present, due to COVID-19 pandemic, we were required to limit our mobility significantly. All these enforced changes in personal and professional lives have resulted in a huge surge of digital footprints. The present-day ICT services, both online and offline are not in general designed to include conducive options for digital legacy management. We are now creating new contents, opening new digital service accounts, and conducting online financial transactions at a much higher rate. Hence, another disturbing trend we have observed to be on the rise is the potential identity theft and targeted blackmailing through exploiting the deceased users' accounts.

## 4.2. Digital Afterlife and Religion- Mourning of The Deceased Users Online

The validity of various online funerals and mourning sessions through different religious lenses has also initiated a lot of confusions and conversation within the space of religion, society, and technology. Funerals and last rituals performed over Zoom, FaceTime, or Skype became a new normal during this period of COVID-19 (Observer Report, 2021), It has been observed that various online groups and online chats have been formed voluntarily by the families and friends of a deceased person to mourn and perform religious activities during the pandemic (Bray, 2021). Despite the lack of internet or electricity connectivity in parts of India, Indonesia and The Philippines families of the deceased tried to perform online religious activities to pray for the departed soul (Bray, 2021).

During COVID-19, we have observed a shift towards online platforms to mourn, even as social distancing is coming at an ease. However, expectations of privacy and ownership of such online events (and the related content) in the long term is unclear. According to Bray, "...the pandemic is transforming the global "death care" market, which consists of funeral, burial, and internment services and which, in 2020, was estimated to have a market size of $106.3 billion" (Bray, 2021). Startups are beginning to capitalize on the "end of life sector, Guruji on Demand (Indian Hindu ritual service startup) charges $339 to take care of everything (revenue jumped 219% from the first quarter of 2020 to the end of September) (Bray, 2021). Therefore, unknowingly the demand for all online services for religious rituals for a deceased increased. For instance, Fu Shou Yuan International Group (China's largest funeral provider) offers services that allow mourners to "virtually clean" the graves of their loved ones and upload virtual offerings (Bray, 2021). The demand for end of life (shuūkatsu) services has shown in a survey in 2020 that 28% of Japanese had signed up with a funeral company (Bray, 2021).

## 4.3. New Services with Limited Digital Afterlife Policies

According to our initial research, few of the new and leading digital communication and professional services have Digital Afterlife policies for its users to manage their post-mortem data, both nationally and internationally. For instance, in February 2020, the only existing Google policy directly related to the user's Digital Afterlife is the Inactive Account Manager (IAM), which allows users to determine what happens to their accounts after a certain period of inactivity (Google, "Inactive Account Manager", 2021). In addition, other start-ups such as *GoodTrust,* "Provide information about the account, and upload documents like proof that the person has passed away and create necessary online Power of Attorneys (GoodTrust, 2021). *SafeBeyond*, uses Amazon's Cloud technology to store sensitive information in "vault-like" servers that cannot be accessed by third parties (SafeBeyond, 2021), *Afternote*, provides users with the opportunity to digitally store their life story, leave messages to their loved ones and record their last will (Afternote, 2021). *Eter9*, social network that uses Artificial Intelligence to create a virtual being (deceased) that publishes, comments, and interacts with you intelligently (Eter9, 2021). Nevertheless, in most cases, according to our observation, people have limited knowledge about the everlasting consequences of such massive digital presence. Majority of the users in the Global South are neither aware of the concept of Digital Afterlife nor the traditional challenges it poses (as mentioned above), making them further vulnerable online.

Moreover, we are observing a big spike of the number of deceased users' social media and other digital service accounts in our focused counties due to COVID-19. We aim to closely study some of these accounts closely to see how those are managed (or left behind or exploited) in the coming days. Such knowledge would help us to better understand our societal vulnerability in the absence of proper awareness options.

# 5. FUTURE WORK

## 5.1. Raising Awareness About Digital Afterlife

According to our research, lack of understanding about the basic concepts of Digital Afterlife is a major obstacle in Global South. People are still unaware about the negative impacts (e.g., copyright infringement, privacy violation, identity theft, loss of private data, etc.) of a poorly managed digital account of a deceased person. Digital Afterlife specific factors have to be included in the ICT policies with relative ease alongside the traditional perspectives of online copyright and intellectual property rights in in India, Indonesia and The Philippines.

## 5.2. Digital Afterlife and Religion

In the long history of human society, afterlife has played a prominent role in shaping up significant socio-religious events and activities. During this period of global pandemic, the situation has become even more challenging. Our initial findings showed that people are forced to transition to digital platforms to perform religious rituals and mourning for the deceased. It also opened the need for new interpretations of afterlife in relation to the ongoing digital presence of any human being. Through our research, we plan to engage with the religious experts and general population to know more about their perceptions and interpretations related to this new phenomenon. Our findings can shed some lights on the entangled worlds of social media and religion.

## 5.3. Responsible and Inclusive Service Providers

Our aim is to ensure that the disparity between knowledge and services are improved in both Global South and Global North. We need to understand what the society is aware of and whether we need to address the Digital Afterlife concept from a cultural or religious point of view or implement policies. Moreover, creating inclusive ICT policy framework, which without blindly following the western solutions, can recognize the Digital Afterlife specific challenges and localized solutions. The proposal is to work on raising awareness about Digital Afterlife by creating dialogues among different stakeholders, in Public Interest Technology forum. To mitigate the disconnection and disparity between Global North and Global South in aspect of a global pandemic, COVID-19 and Digital Afterlife. We eventually envision proposing a set of recommendations, based on our findings, for developing a more inclusive and equitable digital space in this pandemic-stricken world.

# REFERENCES AND CITATIONS